# Simplified Gyroscopic Treatment

M. N. Tarabishy

goodsamt@gmail.com

**Abstract:**

Gyroscopic motion explanation in texts is relatively long and requires reasonable level of comfort with the mathematical tools used.  On the other hand, popular explanation outside academic courses does not explain the phenomenon and only describes it leaving many to think that it is so weird that it defies physics.  In this paper we offer a simplified and mathematically sound explanation that can be used in either setting.

## 1. Introduction:

One of the most interesting concepts in classical physics is the motion of the gyroscope or in short, the gyro.  It is an object that spins around its main axis.

Understanding this phenomenon is not only important for earth science, astronomy, aerospace applications, stabilization of vessels and navigation but also for training of pilots on how gyroscopic instruments work.  Popular treatment of the subject usually describes the gyro behavior without providing explanation.

One of the best popular demonstrations is a video by Derek Muller [1].  He uses the bicycle wheel experiment where the spinning wheel appears to defy gravity and starts to "precess" instead of falling down as sketched in figure 1, where on the left, the non spinning wheel hangs from the string but a spinning wheel balances its weight effect while precessing.

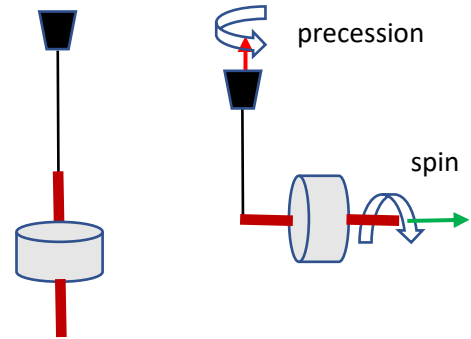

**Figure 1.** Bicycle wheel gyro experiment.

The gyroscopic phenomenon is explained well in physics books from introductory texts like Hibbler [2], to more advanced texts like Ginsberg [3], but the explanation can be a bit hard for the uninitiated as it takes some level of comfort with the mathematical tools used. Therefore, we suggest the use of a simpler explanation that is mathematically sound.

## 2. The Simplified Approach:

### 2-1. Free Gyroscope:

In the case of free spinning wheel as shown in figure 2, the sum of external moments around its center of mass, point G, is zero:

$$\boldsymbol{M}_G = \boldsymbol{0} \quad (1)$$

The angular momentum **H** of an object is inertia matrix J times the angular veocity vector **ω**.  But for the gyro that is spininng with angular velocity ω around its principal axis of symmetry Z:

$$\boldsymbol{H}_G = J\boldsymbol{\omega} = \boldsymbol{H}_z = J_z \cdot \boldsymbol{\omega} \quad (2)$$

With:  $J_z$ is the moment of inertia around Z.

Using Euler equation that states: Resultant moment affecting an object = time change of its angular momentum. In this case:

$$M_G = \frac{dH_G}{dt} = 0 \Rightarrow \quad (3)$$
$$H_G = constant \quad (4)$$

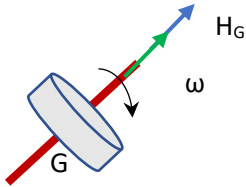

**Figure 2.** Spinning free wheel.

Meaning that the angular momentum vector is constant in an inertial frame.

This concept is sometimes referred to as "rigidity in space" and is used as a basis for gyro compass and aircraft instruments like Heading Indicator and Attitude Indicator (artificial horizon).

The physical realization of the free gyro is to mount the spinning wheel in gimbals like figure 3. The disk spins in the inner gimbal that can rotates relative to the outer one which is free to rotate around its base, therefore, the disk maintains its orientation in an inertial frame. However, friction and imbalance will usually cause a slow drift.

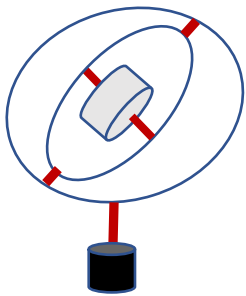

**Figure 3.** Free gyro.

## 2-2. Precession Due to Moment on Gyro Axis:

Using a spinning disk with its axis along the Z-axis as shown in figure 4.

The spin velocity is indicated by vector **ω₁**, and the disk angular momentum by **H_G** with its tip at point B. When force **F** acts on point B at a distance d, it produces a moment **M_G** along the Y-axis causing point B to move. Using Resal Theorem [4], the speed of point B, **V_B**, is the time change of **H_G** and is equal to the resultant moment **M_G** of the forces around the center G, so:

$$GB = d = H_G = H_z = \omega_1 J_z \quad (5)$$
$$\frac{dH_G}{dt} = M_G = d \times F$$
$$\frac{dGB}{dt} = V_B = M_G = \omega_2 \times H_z \quad (6)$$

So, in the simple case shown, we get:

$$\omega_2 = \frac{M_G}{J_Z \cdot \omega_1} \quad (7)$$

Note: The direction of axis movement is at 90° to the force in the direction of spinning.

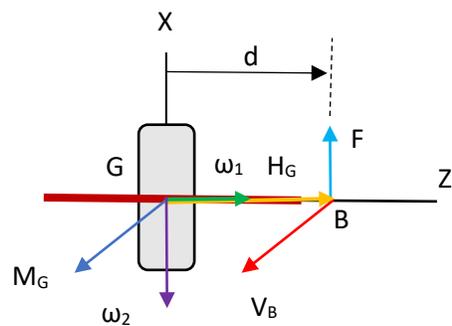

**Figure 4.** Analysis of disk precession.

In summary, when a torque M_G affects a spinning object, it starts to turn. The tip of the angular momentum vector Hz acquires a velocity V_B. That velocity is its change in time dHz/dt which IS EQUAL to the torque $V_B = M_G$. The velocity of the tip is also equal to the vector (angular momentum $J_z \omega_1$) times its turn angular velocity $\omega_2$ (precession), $M_G = J_z \omega_1 \omega_2$.

This property is used in aircraft Turn Coordinator to measure the angular velocity.

## 3. The General Derivation:

### 3-1. Precession Due to Moment on Gyro Axis:

In the following, and for comparison, we examine textbooks approach.

The angular velocity vector in a body frame {x,y,z} is:

$$\boldsymbol{\omega} = \omega_x \hat{\imath} + \omega_y \hat{\jmath} + \omega_z \hat{k} \qquad (8)$$

We can use Euler angles to describe the orientation of the object as shown in figure 5:

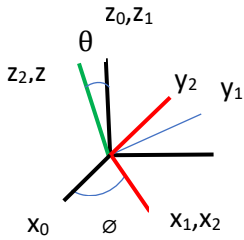

**Figure 5.** Euler angles and corresponding frames.

We start with the inertial frame {$x_0,y_0,z_0$}

Rotate around $z_0$ by angle $\varnothing$ resulting in frame {$x_1,y_1,z_1$}.

Followed by rotation around $x_1$ by an angle θ resulting in frame {$x_2,y_2,z_2$}. This frame has the property that the inertia of the axisymmetric body is fixed in it.

Spinning around $z_2$ by angle ψ will give the body frame {x,y,z}.

Then,

$$\boldsymbol{\omega} = \dot{\phi}\widehat{k_0} + \dot{\theta}\widehat{\imath_2} + \dot{\psi}\hat{k} \qquad (9)$$

Or,

$$\begin{aligned}\boldsymbol{\omega} = &\left[\dot{\theta}\cos(\psi) + \dot{\phi}\sin(\theta)\sin(\psi)\right]\hat{\imath} \\ &+ \left[-\dot{\theta}\sin(\psi) + \dot{\phi}\sin(\theta)\cos(\psi)\right]\hat{\jmath} \\ &+ \left[\dot{\phi}\cos(\theta) + \dot{\psi}\right]\hat{k}\end{aligned} \qquad (10)$$

However, we can choose frame {$x_2,y_2,z_2$} that doesn't spin with the body and has an angular velocity $\boldsymbol{\Omega}$.

$$\boldsymbol{\omega} = \boldsymbol{\Omega} + \dot{\psi}\hat{k} \qquad (11)$$

$$\boldsymbol{\Omega} = \dot{\theta}\widehat{\imath_2} + \dot{\phi}\sin(\theta)\widehat{\jmath_2} + \dot{\phi}\cos(\theta)\widehat{k_2} \qquad (12)$$

Using modified Euler equation:

$$\boldsymbol{M} = \frac{d\boldsymbol{H}}{dt} + \boldsymbol{\Omega} \times \boldsymbol{H} \qquad (13)$$

$$M_x = \dot{H}_x + (J_z\omega_z\Omega_y - J_y\omega_y\Omega_z) \qquad (14)$$
$$M_y = \dot{H}_y + (J_x\omega_x\Omega_z - J_z\omega_z\Omega_x) \qquad (15)$$
$$M_z = \dot{H}_z + (J_y\omega_y\Omega_x - J_x\omega_x\Omega_y) \qquad (16)$$

Or,

$$\begin{aligned}M_x = &J_x\ddot{\theta} + J_z\dot{\phi}\sin(\theta)\left(\dot{\phi}\cos(\theta) + \dot{\psi}\right) \\ &- J_y\dot{\phi}^2\sin(\theta)\cos(\theta)\end{aligned} \qquad (17)$$

$$\begin{aligned}M_y = &J_y[\ddot{\phi}\sin(\theta) + \dot{\theta}\dot{\phi}\cos(\theta)] \\ &+ J_x\dot{\theta}\dot{\phi}\cos(\theta) - J_z\dot{\theta}[\dot{\phi}\cos(\theta) + \dot{\psi}]\end{aligned} \qquad (18)$$

$$\begin{aligned}M_z = &J_z[\ddot{\phi}\cos(\theta) + \ddot{\psi} - \dot{\phi}\dot{\theta}\sin(\theta)] \\ &+ J_y\dot{\theta}\dot{\phi}\sin(\theta) - J_x\dot{\theta}\dot{\phi}\sin(\theta)\end{aligned} \qquad (19)$$

The relation between the Euler rates and angular velocity vector components is

$$\begin{bmatrix}\omega_x \\ \omega_y \\ \omega_z\end{bmatrix} = \begin{bmatrix}s\theta s\psi & c\psi & 0 \\ s\theta c\psi & -s\psi & 0 \\ c\theta & 0 & 1\end{bmatrix}\begin{bmatrix}\dot{\phi} \\ \dot{\theta} \\ \dot{\psi}\end{bmatrix} \qquad (20)$$

For axisymmetric object with steady precession,

$$J_x = J_y = J, \quad \theta, \dot{\phi}, \dot{\psi} = const. \quad (21)$$

Then,

$$M_x = J_z \dot{\phi} \sin(\theta)\left(\dot{\phi}\cos(\theta)+\dot{\psi}\right) - J\dot{\phi}^2 \sin(\theta)\cos(\theta) \quad (22)$$

And

$$M_y = M_z = 0 \quad (23)$$

Or,

$$M_x = \dot{\phi}\sin(\theta)[J_z \omega_z - J\dot{\phi}\cos(\theta)] \quad (24)$$

When $\omega_z \gg \Omega_z$, then,

$$M_x = \dot{\phi}\sin(\theta)J_z\omega_z \quad (25)$$

When $\theta = 90°$, then,

$$M_x = J_z \dot{\phi}\dot{\psi} \quad (26)$$

Which is the same as equation (7)

As a result of the luni-solar effect on the earth bulge, the earth has a precession $\dot{\phi}$ with a period of about 25,800 years.

**4-2. Free Gyro:**

Starting with Euler equations:

$$M_x = \dot{H}_x + (J_z\omega_z\omega_y - J_y\omega_y\omega_z) \quad (27)$$
$$M_y = \dot{H}_y + (J_x\omega_x\omega_z - J_z\omega_z\omega_x) \quad (28)$$
$$M_z = \dot{H}_z + (J_y\omega_y\omega_x - J_x\omega_x\omega_y) \quad (29)$$

For torque free axisymmetric object:

$$J_x = J_y = J, \quad M_{x,y,z} = 0 \quad (30)$$

Therefore, eqn. (29) gives:

$$\dot{H}_z = 0 \Rightarrow \omega_z = const. \quad (31)$$

Eqn. (27) gives:

$$\dot{\omega}_x + \omega_y\omega_z \frac{(J_z - J)}{J} \quad (32)$$

So, eqns. (27) and (28) can be written as:

$$\dot{\omega}_x + \gamma\,\omega_y = 0 \quad (33)$$
$$\dot{\omega}_y - \gamma\,\omega_x = 0 \quad (34)$$

Multiplying by $\omega_x$, $\omega_y$ then adding

$$\dot{\omega}_x^2 + \dot{\omega}_y^2 = \dot{\omega}_{xy}^2 = 0 \Rightarrow \omega_{xy} = const. \quad (35)$$

And,

$$H_{xy} = J\,\omega_{xy} = const., \quad \omega = const. \quad (36)$$

So, the projections of $\omega$, and H on the x-y plane are along the same line. The three vectors: **z, H, ω** are coplanar.

Defining the projection of the angular velocity vector on the x-y plane [5]:

$$\omega_{xy} = \omega_x + i\,\omega_y \quad (37)$$

Then, equations (33), (34) give:

$$\dot{\omega}_{xy} - i\,\gamma\,\omega_{xy} = 0 \quad (38)$$

With a solution:

$$\omega_{xy} = \omega_{xy}(0)\,e^{i\gamma t} \quad (39)$$

So, $\omega_{xy}$ rotates in the x-y plane with angular speed $\gamma$.

$$\gamma = \omega_z \frac{J_z - J}{J} \quad (40)$$

Using figure 6, we can write:

$$H_{xy} = H\sin(\theta),\; H_z = H\cos(\theta) \quad (41)$$

$$\omega_{xy} = \omega\sin(\alpha),\; \omega_z = \omega\cos(\alpha) \quad (42)$$

Then,

$$\tan(\theta) = \frac{J}{J_z}\tan(\alpha) \quad (43)$$

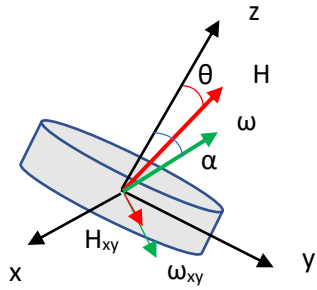

**Figure 6.** Angular velocity and momentum vectors.

Using Euler angles for ω at ψ = 0 :

$$\boldsymbol{\omega} = [\dot{\phi} \sin(\theta)]\hat{j} + [\dot{\phi} \cos(\theta) + \dot{\psi}]\hat{k} \quad (44)$$

Gives:

$$\dot{\theta} = 0 \quad (45)$$

$$\dot{\phi} = \frac{J_z}{J \cos(\theta)} \omega_z \quad (46)$$

$$\dot{\psi} = \frac{J - J_z}{J} \omega_z = -\gamma \quad (47)$$

Another important a relation:

$$\dot{\psi} = \frac{J - J_z}{J_z} \dot{\phi} \cos(\theta) \quad (48)$$

For earth, $J_z/J$ =1.003284, α ≈ θ = 0.2"

$\omega_z$ =2π/23h 56' 4.1"= 7.292E-5 rad/s

$\dot{\phi}$ = 7.316E-5 rad/s, ϒ = -2.395E-7 rad/s

Precession period for $\dot{\phi}$ is 0.994 days.

The period for ϒ is 304 days "Chandler Wobble". It is different than the measured 434 days due to nonrigidity.

## 4. Discussion and Conclusions:

In this paper, we have done a simple but mathematically sound explanation of gyroscopic motion, and we also done detailed academic treatment for comparison. The detailed treatment rewards you with more details and deeper understanding not possible with the simplified treatment, however, the simplified treatment is much shorter and simpler and gives a more accessible way to a large audience.

## 5. References:

[1] Veritasium: Gyroscopic Precession

https://youtu.be/ty9QSiVC2g0

[2] Hibbler, R. C., Engineering Mechanics: Dynamics, Prentice Hall, New Jersey, 1998.

[3] Ginsberg, J. H., Advanced Engineering Dynamics, 2nd ed., Cambridge University Press, 1995.

[4] Targ, S., Theoretical Mechanics: A Short Course, Mir publishers, Moscow, 1988.

[5] Meirovitch, L., Introduction to Dynamics and Control, John Wiley & Sons, NY. 1985.

[6] D'Souza, A. F, & Garg, V. K., Advanced Dynamics, Prentice Hall, New Jersey, 1984.

[7] Greenwood, D. T., Principles of Dynamics, Prentice Hall, New Jersey, 1988.

[8] Moon, F. C., Applied Dynamics, John Wiley & Sons, NY. 1998.

[9] Precession Intuitively Explained

https://www.frontiersin.org/articles/10.3389/fphy.2019.00005/full